\documentclass[11pt,a4paper]{article}
\pdfoutput=1

\usepackage{jheppub_kim}
\usepackage{rotating}
\usepackage{graphicx,epsfig}
\usepackage{amsmath}
\usepackage {amssymb}
\usepackage{subfigure}

\usepackage{relsize}

\def\beq{\begin{equation}}
\def\eeq{\end{equation}}
\def\bea{\begin{eqnarray}}
\def\eea{\end{eqnarray}}

\begin{document}

\title{New observational constraints on $f(T)$ gravity from cosmic chronometers}

\author[a]{Rafael C. Nunes}

\author[b]{Supriya Pan}

\author[c,d,e]{Emmanuel N. Saridakis}

 \affiliation[a]{Departamento de F\'isica, Universidade Federal de Juiz de Fora,
36036-330,
Juiz de
Fora, MG, Brazil}

 \affiliation[b]{Department of Physical Sciences, Indian Institute of Science Education
and Research -- Kolkata, Mohanpur -- 741246, West Bengal, India}

\affiliation[c]{Instituto de F\'{\i}sica, Pontificia
Universidad de Cat\'olica de Valpara\'{\i}so, Casilla 4950,
Valpara\'{\i}so, Chile}

\affiliation[d]{CASPER, Physics Department, Baylor University, Waco, TX 76798-7310, USA}

\affiliation[e]{Physics Division,
National Technical University of Athens, 15780 Zografou Campus,
Athens, Greece}

\emailAdd{nunes@ecm.ub.edu}
\emailAdd{span@iiserkol.ac.in}
\emailAdd{Emmanuel$_-$Saridakis@baylor.edu}

\abstract{
We use  the local value of the Hubble constant recently measured with 2.4$\%$ precision,
as well as the latest compilation of  cosmic chronometers data, together with standard
probes such as Supernovae Type Ia   and Baryon Acoustic Oscillation distance
measurements, in order to impose constraints on the viable and most used $f(T)$ gravity
models,  where $T$ is the torsion scalar in teleparallel gravity. In particular, 
we 
consider three $f(T)$ models with two parameters, out of which one is independent,  and 
we 
quantify 
their deviation from $\Lambda$CDM cosmology through
a sole parameter. Our analysis reveals that for one of the models a small but non-zero
deviation from $\Lambda$CDM cosmology is slightly favored, while for the other models  the
best fit  is very close to  $\Lambda$CDM scenario. Clearly, $f(T)$ gravity
is consistent with observations, and it can serve as a candidate for modified
gravity.
}


\maketitle

\section{Introduction}

According to a large amount of observational data the universe is currently exhibiting an
accelerating expansion, while according to theoretical arguments and observational
indications at early times the universe should had also experienced a phase of
accelerated expansion called inflation. There are two main directions that one could
follow in order to provide a description. The first is to remain in the framework of
general relativity and introduce new, exotic components in the universe content, such as
the inflaton field(s) \cite{Guth:1980zm,Linde:1981mu} and the dark energy sector
\cite{Copeland:2006wr,Amendola,Cai:2009zp}. The second way is to construct gravitational
modifications in such a way that the additional gravitational degree(s) of 
freedom could
drive the accelerating expansion \cite{Capozziello:2011et}.

Concerning modified gravity, the large majority of works start from the standard
gravitational description, i.e. from its curvature formulation, and extend the
Einstein-Hilbert action, as for instance in the  $F(R)$ gravity
\cite{Nojiri:2010wj,DeFelice:2010aj}, in
Gauss-Bonnet and $f(G)$ gravity \cite{Nojiri:2005jg,DeFelice:2008wz}, in Lovelock gravity
\cite{Lovelock:1971yv,Deruelle:1989fj}, in Ho\v{r}ava-Lifshitz gravity
\cite{Horava:2008ih}, in massive gravity \cite{deRham:2014zqa} etc. However, one can
equally well construct gravitational modifications  starting from the
torsion-based formulation, and specifically from  the Teleparallel Equivalent of General
Relativity (TEGR)
\cite{Unzicker:2005in,TEGR,TEGR22,Hayashi:1979qx,JGPereira,Maluf:2013gaa}. Since in this
theory
the Lagrangian is the torsion scalar $T$, the simplest modification is
the $f(T)$ gravity \cite{Bengochea:2008gz,Linder:2010py} (see
\cite{Cai:2015emx} for a review), a theory that is different from $f(R)$ gravity despite
the fact that TEGR is completely equivalent with general relativity at the level of
equations. That is why $f(T)$ gravity and its cosmological applications have gained a lot
of interest in the literature, both for   early-time
\cite{Ferraro:2006jd,Ferraro:2008ey,Bamba:2014zra,Bamba:2016gbu,Bamba:2016wjm}
and late-time
\cite{Chen:2010va,Wu:2010xk,Tsyba:2010ji,Yang:2010hw,Bamba:2010iw,
Myrzakulov:2010tc,Bamba:2010wb,Dent:2011zz,Cai:2011tc,Sharif:2011bi,Wei:2011mq,Wu:2011xa,
Karami:2011np,
Jamil:2011mc,Karami:2012if,Setare:2012vs,Dong:2012en,Tamanini:2012hg,Daouda:2012wt,
Banijamali:2012nx,Darabi:2012zh,Liu:2012fk,Setare:2012ry,Chattopadhyay:2012eu,
Bamba:2012ka, 
Setare:2013xh,Li:2013xea,Bamba:2013fta,Basilakos:2013rua,Qi:2014yxa,Nashed:2014baa,
Darabi:2014dla,
Nashed:2015pda,Fayaz:2015yka,Abedi:2015cya,Krssak:2015oua,Pan:2016jli,
Paliathanasis:2016vsw}
 descriptions. Finally, one can construct various extensions of $f(T)$ gravity 
and
investigate their interesting cosmological implications
\cite{Geng:2011aj,Xu:2012jf,Otalora:2013tba, deHaro:2012zt, 
Otalora:2013dsa,Kofinas:2014owa,Harko:2014sja,
Harko:2014aja,Jawad:2015szv,Skugoreva:2014ena,Gonzalez:2015sha,Bahamonde:2015zma,
Bahamonde:2015hza,Sadjadi:2015fca,Skugoreva:2016bck,Otalora:2016dxe, Junior:2015kia}.

A crucial question in every modified gravity that incorporates an unknown function, is
what are the forms of this function and moreover what are the values of the involved
parameters. As long as one excludes forms and parameter regions that lead to
contradictions or theoretical disadvantages, the main tool he has in order to further
constrain the remaining classes is to use observational data. In the case of $f(T)$
gravity this has been done using cosmological observations from Supernovae type Ia,
cosmic
microwave background and baryonic acoustic oscillations
\cite{Wu:2010mn,Nesseris:2013jea,Capozziello:2015rda,Basilakos:2016xob}
or solar system data \cite{Iorio:2012cm,Iorio:2015rla,Farrugia:2016xcw}.

In this work we are interested in providing updated constraints on various $f(T)$ gravity
models, using the very recently released data  \cite{riess} .  In particular,
we use the latest astronomical data sets: (1)- Hubble parameter measurements from the
differential evolution of cosmic chronometers (CC) $+$ the latest released local value
of the Hubble parameter ($H_0$) at 2.4\% precision; (2)- Type Ia Supernovae sample from
Union 2.1 compilation containing 580 data points, and finally (3)- six baryon acoustic
oscillation data. The joint analysis of the above data sets provides new constraints on
$f(T)$ models.  The manuscript is organized  as follows: In Section \ref{fTcosmology} we  
briefly 
review $f(T)$ gravity and cosmology, focusing on
the three viable and thus most-used models. In Section \ref{data} we present
the latest data sets for our analysis, whereas in Section \ref{sec:results} we use them
in order to extract  constraints on various quantities.  Finally, in Section
\ref{Conclusions} we   summarize our results.

\section{$f(T)$ gravity and cosmology}
\label{fTcosmology}

In this section we briefly review $f(T)$ gravity and we apply it in a cosmological
framework. Then we examine three specific $f(T)$ models, which are the viable ones
amongst the variety of $f(T)$ scenarios, since they pass the basic observational tests.

\subsection{$f(T)$ gravity}

In $f(T)$ gravity, and similarly to all torsional formulations, we use the
vierbein fields $e^\mu_A$, which form an orthonormal base on the tangent space
at each manifold point $x^{\mu}$. The metric then reads as
$g_{\mu\nu}=\eta_{A B} e^A_\mu e^B_\nu$ (in this manuscript greek indices
and Latin indices span respectively the coordinate and tangent spaces). Moreover,
instead of the torsionless Levi-Civita connection we use the curvatureless
Weitzenb{\"{o}}ck one $\overset{\mathbf{w}}{\Gamma}^\lambda_{\nu\mu}\equiv e^\lambda_A\:
\partial_\mu e^A_\nu$  \cite{JGPereira}, and hence the gravitational field is
described by the torsion tensor
\begin{equation}
T^\rho_{\verb| |\mu\nu} \equiv e^\rho_A
\left( \partial_\mu e^A_\nu - \partial_\nu e^A_\mu \right).
\end{equation}
The Lagrangian of teleparallel equivalent of general relativity, i.e. the torsion scalar
$T$, is
constructed by contractions of the torsion tensor as \cite{JGPereira}
\begin{equation}
\label{Tscalar}
T\equiv\frac{1}{4}
T^{\rho \mu \nu}
T_{\rho \mu \nu}
+\frac{1}{2}T^{\rho \mu \nu }T_{\nu \mu\rho}
-T_{\rho \mu}{}^{\rho }T^{\nu\mu}{}_{\nu}.
\end{equation}
Inspired by the $f(R)$ extensions of general relativity  we can  extend $T$ to a function
$T+f(T)$, constructing the action of $f(T)$ gravity \cite{Bengochea:2008gz,Linder:2010py}:
\begin{eqnarray}
\label{actionbasic}
 {\mathcal S} = \frac{1}{16\pi G}\int d^4x e \left[T+f(T)\right],
\end{eqnarray}
with $e = \text{det}(e_{\mu}^A) = \sqrt{-g}$ and $G$ the gravitational
constant (we have imposed units where the light speed is equal to 1). Note that
TEGR and thus general
relativity is restored when $f(T)=0$, whereas for $f(T)=const.$, we recover
general relativity with a cosmological constant.

\subsection{$f(T)$ cosmology}

Let us apply $f(T)$ gravity in a cosmological framework. Firstly, we need to incorporate
the matter and the radiation sectors, and thus the total
action is written as
\begin{eqnarray}
\label{action11mat}
 {\mathcal S}_{tot} = \frac{1}{16\pi G }\int d^4x e
\left[T+f(T)\right] \, + {\mathcal S}_m+ {\mathcal S}_r   ,
\end{eqnarray}
with the matter and radiation Lagrangians  assumed to correspond
to perfect fluids with energy densities $\rho_m$, $\rho_r$ and pressures $P_m$, $P_r$
respectively.

Variation of the action (\ref{action11mat}) with
respect to the vierbeins provides the field equations as
\begin{eqnarray}
\label{eom}
&&\!\!\!\!\!\!\!\!\!\!\!\!\!\!\!
e^{-1}\partial_{\mu}(ee_A^{\rho}S_{\rho}{}^{\mu\nu})[1+f_{T}]
 +
e_A^{\rho}S_{\rho}{}^{\mu\nu}\partial_{\mu}({T})f_{TT}
-[1+f_{T}]e_{A}^{\lambda}T^{\rho}{}_{\mu\lambda}S_{\rho}{}^{\nu\mu}+\frac{1}{4} e_ { A
} ^ {
\nu
}[T+f({T})] \nonumber \\
&&= 4\pi Ge_{A}^{\rho}
\left[{\mathcal{T}^{(m)}}_{\rho}{}^{\nu}+{\mathcal{T}^{(r)}}_{\rho}{}^{\nu}\right],
\end{eqnarray}
with $f_{T}=\partial f/\partial T$, $f_{TT}=\partial^{2} f/\partial T^{2}$,
and where ${\mathcal{T}^{(m)}}_{\rho}{}^{\nu}$ and  ${\mathcal{T}^{(r)}}_{\rho}{}^{\nu}$
are  the matter and radiation energy-momentum tensors respectively.

As a next step we focus on homogeneous and isotropic geometry, considering the
usual choice for the vierbiens, namely
\begin{equation}
\label{weproudlyuse}
e_{\mu}^A={\rm
diag}(1,a,a,a),
\end{equation}
which corresponds to a flat Friedmann-Robertson-Walker (FRW) background
 metric of the form
\begin{equation}
ds^2= dt^2-a^2(t)\,\delta_{ij} dx^i dx^j,
\end{equation}
where $a(t)$ is the scale factor.

Inserting the vierbein   (\ref{weproudlyuse}) into the field equations
(\ref{eom}) we acquire the Friedmann equations as
\begin{eqnarray}\label{background1}
&&H^2= \frac{8\pi G}{3}(\rho_m+\rho_r)
-\frac{f}{6}+\frac{Tf_T}{3}\\\label{background2}
&&\dot{H}=-\frac{4\pi G(\rho_m+P_m+\rho_r+P_r)}{1+f_{T}+2Tf_{TT}},
\end{eqnarray}
with $H\equiv\dot{a}/a$ the Hubble parameter, and where we use dots to denote derivatives
with respect to $t$. In the above relations we have used that
\begin{eqnarray}
\label{TH2}
T=-6H^2,
\end{eqnarray}
which  arises straightforwardly for a FRW
universe through (\ref{Tscalar}).

Observing the form of the first Friedmann equation (\ref{background1}) we deduce that
in $f(T)$ cosmology  we acquire  an effective dark energy sector of
gravitational origin. In particular, we can define the effective
dark energy density and pressure as \cite{Cai:2015emx}:
\begin{eqnarray}
&&\rho_{DE}\equiv\frac{3}{8\pi
G}\left[-\frac{f}{6}+\frac{Tf_T}{3}\right], \label{rhoDDE}\\
\label{pDE}
&&P_{DE}\equiv\frac{1}{16\pi G}\left[\frac{f-f_{T} T
+2T^2f_{TT}}{1+f_{T}+2Tf_{TT}}\right],
\end{eqnarray}
and thus its effective equation-of-state parameter writes as
\begin{eqnarray}
\label{wfT}
 w
=-\frac{f/T-f_{T}+2Tf_{TT}}{\left[1+f_{T}+2Tf_{TT}\right]\left[f/T-2f_{T}
\right] }.
\end{eqnarray}
Note that     $\rho_{DE}$ and $P_{DE}$ defined in (\ref{rhoDDE}),
(\ref{pDE}) obey the usual evolution equation
\begin{eqnarray}
\dot{\rho}_{DE}+3H(\rho_{DE}+P_{DE})=0.
\end{eqnarray}
Finally, the equations close considering the standard matter evolution equation
\begin{eqnarray}
\dot{\rho}_{m}+3H(\rho_{m}+P_{m})=0.
\end{eqnarray}

In this work we are interested in confronting the above Friedmann equations with
observations. Hence, we firstly  define \cite{Nesseris:2013jea}
\begin{eqnarray}
\label{THdef3}
E^{2}(z)\equiv\frac{H^2(z)} {H^2_{0}}=\frac{T(z)}{T_{0}},
\end{eqnarray}
with $T_0\equiv-6H_{0}^{2}$ (in the following we use the subscript ``0'' to denote the
current value of a quantity). Furthermore,  we use the redshift
$z=\frac{a_0}{a}-1$ as the independent variable, and for implicitly we set $a_0=1$.
Therefore, using additionally that
$\rho_{m}=\rho_{m0}(1+z)^{3}$, $\rho_{r}=\rho_{r0}(1+z)^{4}$, we
re-write the first Friedmann equation (\ref{background1}) as \cite{Nesseris:2013jea}
\begin{eqnarray}
\label{Mod1Ez}
E^2(z,{\bf r})=\Omega_{m0}(1+z)^3+\Omega_{r0}(1+z)^4+\Omega_{F0} y(z,{\bf r})
\end{eqnarray}
where
\begin{equation}
\label{LL}
\Omega_{F0}=1-\Omega_{m0}-\Omega_{r0} \;,
\end{equation}
with $\Omega_{i0}=\frac{8\pi G \rho_{i0}}{3H_0^2}$ the corresponding
density parameter at present. In this case the effect of the $f(T)$ modification  is
encoded in the function  $y(z,{\bf r})$ (normalized to
unity at   present time), which depends on $\Omega_{m0},\Omega_{r0}$, and on the
$f(T)$-form parameters $r_1,r_2,...$, namely \cite{Nesseris:2013jea}:
\begin{equation}
\label{distortparam}
 y(z,{\bf r})=\frac{1}{T_0\Omega_{F0}}\left[f-2Tf_T\right].
\end{equation}
We mention that due to  (\ref{TH2}) the additional term (\ref{distortparam})
in the effective
Friedman equation (\ref{Mod1Ez})   is a function of the
Hubble parameter only.

\subsection{Specific viable $f(T)$ models }
\label{manymodels}

In this subsection we review three  specific $f(T)$ models, which are the viable ones
amongst the variety of $f(T)$ scenarios  with two parameters out of which one is
independent, since they pass the basic observational tests \cite{Nesseris:2013jea}.  For
each of them we  calculate the function $y(z,{\bf r})$ through (\ref{distortparam}) and
we quantify the deviation of   $y(z,{\bf r})$ from its (constant)  $\Lambda$CDM value
introducing a distortion parameter $b$.

\begin{enumerate}

\item The power-law model of Bengochea and Ferraro
(hereafter $f_{1}$CDM) \cite{Bengochea:2008gz} is characterized by the form
\begin{equation}
\label{modf1}
f(T)=\alpha (-T)^{b},
\end{equation}
where $\alpha$ and $b$ are the two model parameters.
Inserting this $f(T)$ form
into  Friedmann equation (\ref{background1}) at present, we
acquire
\begin{eqnarray}
\alpha=(6H_0^2)^{1-b}\frac{\Omega_{F0}}{2b-1},
\end{eqnarray}
and moreover (\ref{distortparam})
gives
\begin{equation}
\label{yLL}
y(z,b)=E^{2b}(z,b) \;.
\end{equation}

Clearly, for $b=0$  the present scenario reduces to $\Lambda$CDM cosmology, namely
$T+f(T)=T-2\Lambda$ (with $\Lambda=3\Omega_{F0}H_{0}^{2}$, $\Omega_{F0}=\Omega_{\Lambda
0}$). Finally, we mention that we need  $b<1$ in order to
obtain an accelerating expansion.


\item The Linder model (hereafter $f_{2}$CDM) \cite{Linder:2010py} arises from
\begin{eqnarray}
\label{modf2}
f(T)=\alpha T_{0}(1-e^{-p\sqrt{T/T_{0}}}),
\end{eqnarray}
with $\alpha$ and $p$  the two model parameters. In this case
(\ref{background1})  gives that
\begin{eqnarray}
\alpha=\frac{\Omega_{F0}}{1-(1+p)e^{-p}}\;,
\end{eqnarray}
while
from (\ref{distortparam}) we obtain
\begin{equation}
\label{yLL1}
y(z,p)=\frac{1-(1+pE)e^{-pE}}{1-(1+p)e^{-p}}.
\end{equation}

As we can see,   for $p \rightarrow +\infty$ the present scenario reduces
to  $\Lambda$CDM cosmology. Hence, for convenience we could
replace  $p=1/b$, and thus  (\ref{yLL1}) becomes
\begin{equation}
\label{yLL2}
y(z,b)=\frac{1-(1+\frac{E}{b})e^{-E/b}}{1-(1+\frac{1}{b})e^{-1/b}},
\end{equation}
which indeed tends to unity for
$b \rightarrow 0^{+}$.

 \item Motivated by  exponential $f(R)$ gravity \cite{Linder:2009jz}, Bamba et. al.
introduced the following   $f(T)$ model (hereafter $f_{3}$CDM)  \cite{Bamba:2010wb}:
 \begin{eqnarray}
 \label{modf3}
 f(T)=\alpha T_{0}(1-e^{-pT/T_{0}}),
 \end{eqnarray}
 with $\alpha$ and $p$  the two model parameters.
 In  this case we acquire
 \begin{eqnarray}
 \alpha=\frac{\Omega_{F0}}{1-(1+2p)e^{-p}} \;,
 \end{eqnarray}
 \begin{equation}
 \label{Mod223}
 y(z,p)=\frac{1-(1+2pE^{2})e^{-pE^{2}}}{1-(1+2p)e^{-p}} \;.
 \end{equation}
 Similarly to the previous case, we can re-write the present model using
  $p=1/b$, resulting to
 \begin{equation}
 \label{Mod224}
 y(z,b)=\frac{1-(1+\frac{2E^{2}}{b})e^{-E^{2}/b}}{1-(1+\frac{2}{b})e^{-1/b}} \;.
 \end{equation}
 Hence, we can immediately see that    $f_{3}$CDM model tends to   $\Lambda$CDM
cosmology for $p \rightarrow +\infty$, or
equivalently for $b
 \rightarrow 0^{+}$.

 \end{enumerate}

The above three $f(T)$ forms are the ones that have been used in the
literature of $f(T)$ cosmology, possessing up to two parameters,
out of which one is independent, which have been found to be viable, i.e. they pass
the basic observational tests. There are two more models with two parameters used in the
literature, namely the Bamba et al. logarithmic model \cite{Bamba:2010wb}, with
$f(T)=\alpha T_{0} \sqrt{\frac{T}{qT_{0}}} \ln\left( \frac{qT_{0}}{T}\right )$,  and the
hyperbolic-tangent model   \cite{Wu:2010av}, with $f(T)=\alpha(-T)^{n}\tanh\left(
\frac{T_{0}}{T}\right)$, however since these two models do not possess  $\Lambda$CDM
cosmology as a limiting case and they are in tension with   
observational data  
\cite{Nesseris:2013jea},
we do not consider them in this work. Finally, note that in principle one could
construct $f(T)$ models with more than two parameters, for instance combining the above
simple scenarios, however the appearance of many free parameters would be a significant
disadvantage.

\section{Current Observational Data}
\label{data}

In  this work  we desire to constrain $f(T)$ gravity using observational data obtained by
probes which map the expansion history of the late-time universe (and in particular lying
in the redshift region $z < 2.36$). Our analysis is based on the Hubble parameter
measurements acquired from the cosmic chronometers (CC) technique. On top of that, we
consider the usual observations of Supernovae Type Ia (SNeIa), local Hubble
parameter $H_0$, and Baryon Acoustic Oscillation (BAO). In the following subsections,
we present the different data sets used in our analysis \cite{cc8}.

\subsection{Cosmic chronometer data set and $H_0$}
\label{cc-method-data}

The cosmic chronometers (CC) approach to measure $H(z)$ uses relative ages of the most
massive and passively evolving galaxies in order to measure $dz/dt$, from which one
obtains $H(z)$ \cite{cc1}. The  latest implementation has been described in
\cite{cc2} in  detail, along with the examination of  possible  sources  of uncertainty.
In this work we consider the compilation of Hubble parameter measurements given in
\cite{cc3, cc2}, which contains the latest updated list of $H(z)$ measurements
\cite{cc2,cc3,cc4,cc5,cc6} obtained through the cosmic chronometers approach,
constituting of 30 measurements covering the redshift range $0 < z < 2$. This sample
spans around 10 Gyr of cosmic time. Additionally, in our analysis we  include also
the new local value of $H_0$   measured with a 2.4 $\%$ determination  by \cite{riess},
namely $H_0 = 73.02 \pm 1.79$ km/s/Mpc. These data sets have been recently used to
impose constraints on dynamical dark energy \cite{cc7} and coupled dark energy
\cite{cc8} scenarios.

\subsection{Type Ia Supernovae}
\label{snia-data}

Type Ia Supernovae (SNeIa) are widely used in order to  provide constraints on
dark energy sector, since they serve as ``standard candles'' and thus can offer us a
way to measure cosmic distances. In this work we use the Union 2.1 compilation
\cite{snia3}, which contains 580 SNeIa data in the redshift range $0.015 \leq z \leq
1.41$. The details of the fitting procedure can be found in \cite{cc8}.

\subsection{Baryon Acoustic oscillation}
\label{bao-data}

Baryon acoustic oscillations (BAO) arise from pressure waves at the recombination epoch,
generated by cosmological perturbations in the primeval baryon-photon plasma, and they
appear as distinct peaks in the large angular scales. Hence, they can be used as a
significant cosmological probe for observational analyses. In this work, and similarly
to  \cite{cc8}, we  use  the following 6 BAO data in order to  constrain  the  expansion
history of a given cosmological model:
the  Main Galaxy Sample  of  Data  Release 7  of  Sloan
Digital  Sky  Survey  (SDSS-MGS) \cite{bao2},
the  measurement from the Six Degree Field Galaxy
Survey (6dF) \cite{bao1},  the  LOWZ  and  CMASS  galaxy  samples  of
the Baryon  Oscillation  Spectroscopic  Survey  (namely BOSS-LOWZ  and  BOSS-CMASS)
\cite{bao3}, and the distribution of the LymanForest in BOSS
(BOSS-Ly) \cite{bao4}. The above measurements   and the corresponding
effective redshift are summarized in Table \ref{tab1-bao}.
\begin{table*}[ht]
      \begin{center}
          \begin{tabular}{ccccc}
          \hline
          \hline
           Survey &  $z$     &  Parameter   &  Measurement  & Reference  \\
          \hline
 SDSS-MGS        & 0.10  &  $D_V/r_s$ &  4.47 $\pm$ 0.16   & \cite{bao2} \\
 6dF             & 0.106 & $r_s/D_V$  &  0.327 $\pm$ 0.015 & \cite{bao1} \\
 BOSS-LOWZ       & 0.32  & $D_V/r_s$  &  8.47  $\pm$ 0.17  & \cite{bao3} \\
 BOSS-CMASS      & 0.57  &  $D_V/r_s$ &  13.77  $\pm$ 0.13 & \cite{bao3} \\
 BOSS-$Ly_{\alpha}$& 2.36& $c/(H r_s)$&  9.0    $\pm$ 0.3  & \cite{bao4} \\
 BOSS-$Ly_{\alpha}$& 2.36 & $D_A/r_s$ &  10.08  $\pm$ 0.4  & \cite{bao4} \\
          \hline
          \hline
          \end{tabular}
 	\caption{\label{tab1-bao}The baryon acoustic oscillation (BAO) measurements
used in this work.}
      \end{center}
\end{table*}

\section{Observational Constraints}
\label{sec:results}

In this section we proceed to the main analysis of the present manuscript, and we use the
above observational data in order to impose constraints on the various $f(T)$ models of
subsection \ref{manymodels}. In order to perform the fitting procedure of the involved
free parameters, we use the public code CLASS \cite{Blas:2011rf} along with the public
Monte Carlo code Monte Python \cite{Audren:2012wb}, while as our sampling method we
choose the Metropolis Hastings algorithm. Additionally, for the statistical
analysis we have considered the density parameters for baryons and radiation
at present respectively
as $\Omega_{b0} = 0.05$, $\Omega_{r0}=10^{-5}$, while we use the parameter $\Omega_m$ to
denote
both baryons and cold dark matter, i.e. $\Omega_m= \Omega_{cdm}+ \Omega_b$. In the
following three subsections we perform the fittings for each model separately.

\subsection {$f_{1}$CDM  model:
$f(T)=\alpha (-T)^{b}$}

\begin{figure}[ht]
\centering
\includegraphics[width=0.63\linewidth]{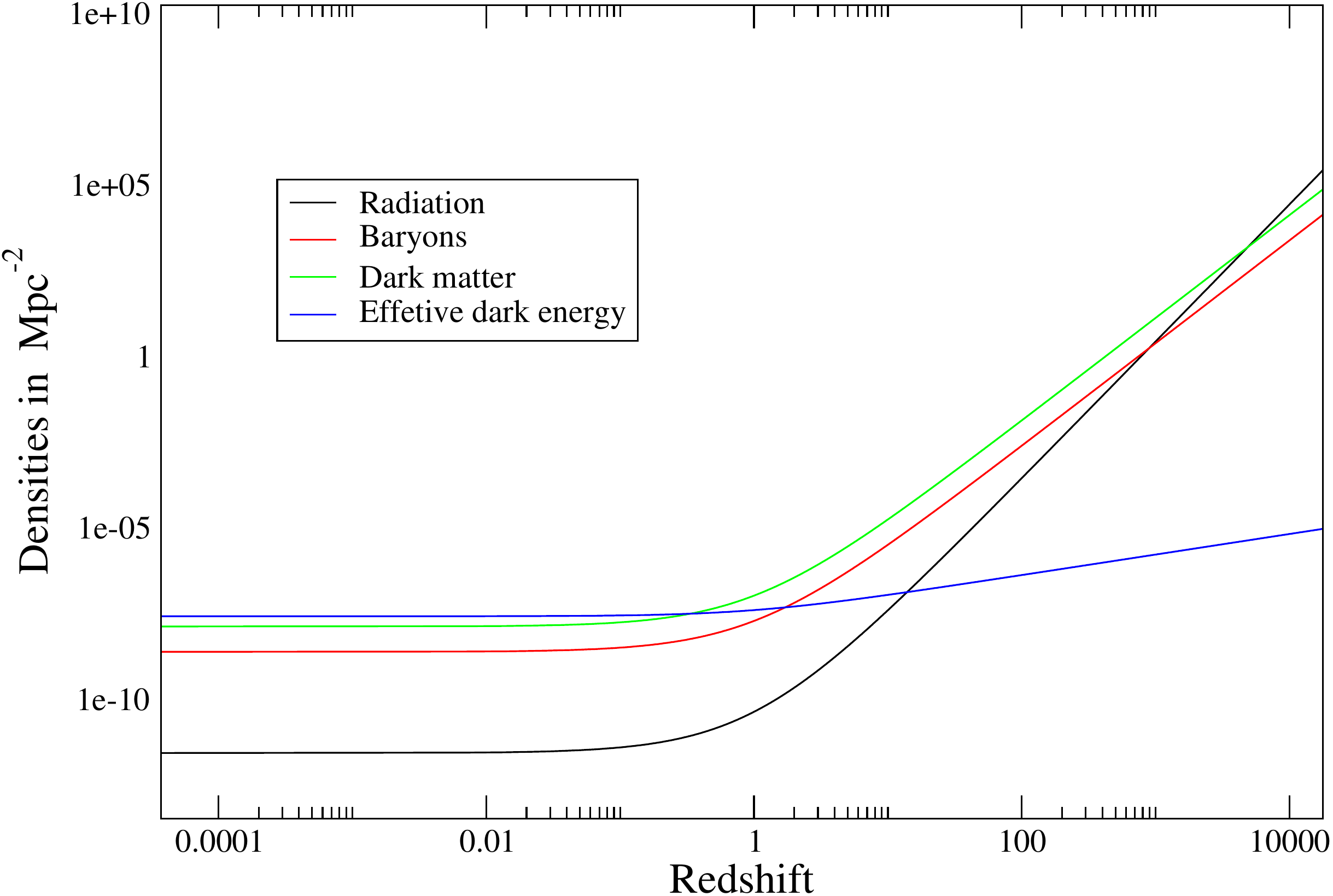}
\caption{{\it{The evolution of various densities in units of $Mpc^{-2}$, multiplied by $8
\pi G/3$, as a function of the redshift, for the $f_1$CDM model: $f(T)=\alpha (-T)^{b}$
of (\ref{modf1}).
 We have considered $b = 0.1$, $h = 0.68$, $\Omega_{b0} = 0.05$, $\Omega_{cdm0} = 0.24$,
   $\Omega_{r0} = 10^{-5}$.}}}
\label{fT1_1}
\end{figure}

For the case of $f_{1}$CDM  model: $f(T)=\alpha (-T)^{b}$ of  (\ref{modf1}), in Fig.
\ref{fT1_1}   we   depict the evolution of the
densities for radiation, baryons, dark matter, and effective dark energy, between the
redshift range $z \in$ [0,10000]. As we can see,
and as expected, the density evolutions are consistent with the thermal history of the
universe, i.e. we obtain   successively the radiation era, the matter era, and at
recent times (the transition is around $z \sim 0.70$) the onset of the dark-energy era
and of cosmic acceleration.

\begin{figure*}
  \includegraphics[width=5.0in, height=5.0in]{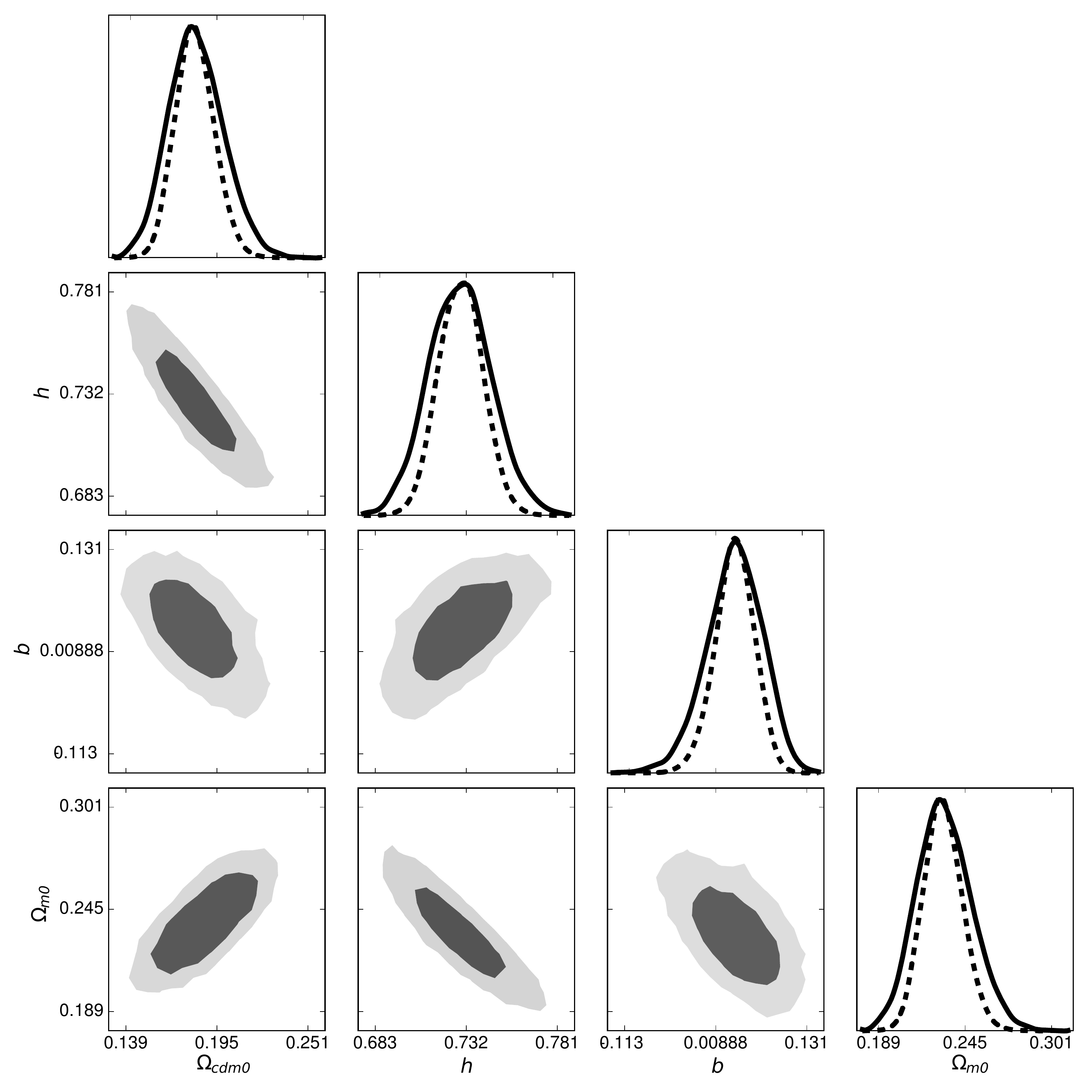}
  \caption{\label{figfT1_cc}
 { \it{ 68.3$\%$ and 95.4$\%$ confidence-level contour plots for various quantities and
for the free parameter $b$, for the  $f_{1}$CDM  model: $f(T)=\alpha (-T)^{b}$ of
(\ref{modf1}), using only $CC$ + $H_0$
observational data. The parameter $\Omega_m$ includes both baryons and cold dark matter,
i.e. $\Omega_m= \Omega_{cdm}+ \Omega_b$,
and \textbf{$h= H_0/100$} km/s/Mpc. Additionally, we present the marginalized
one-dimensional posterior
distribution, where the dashed curve stands for the average likelihood distribution.}}}
\end{figure*}

\begin{figure*}
	\includegraphics[width=5.0in, height=5.0in]{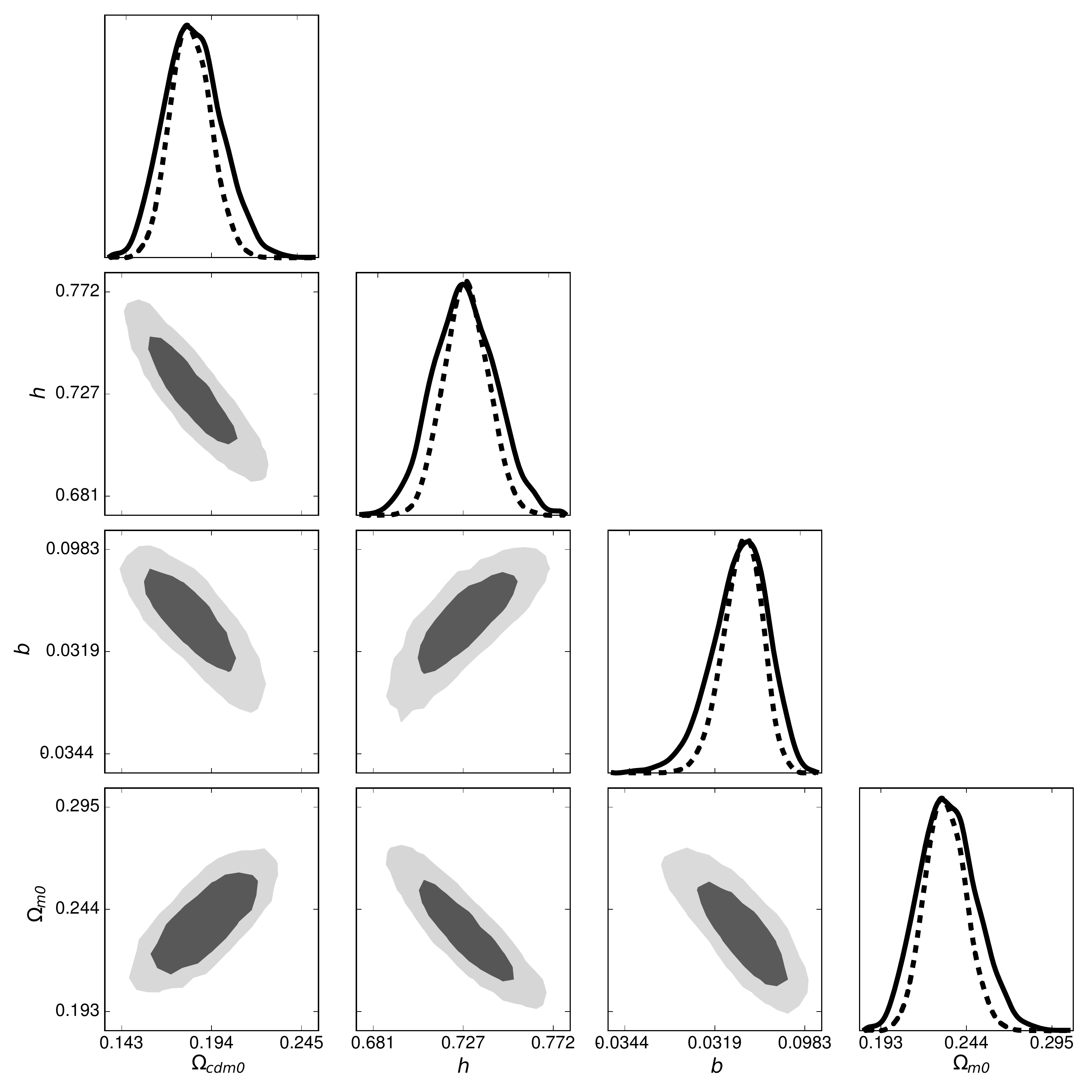}
	\caption{\label{figfT1_joint}
	 { \it{ 68.3$\%$ and 95.4$\%$ confidence-level contour plots for various
quantities and
for the free parameter $b$, for the  $f_{1}$CDM  model: $f(T)=\alpha (-T)^{b}$ of
(\ref{modf1}), using   $CC$ + $H_0$ + SNeIa + BAO
observational data. The parameter $\Omega_m$ includes both baryons and cold dark matter,
i.e. $\Omega_m= \Omega_{cdm}+ \Omega_b$,
and \textbf{$h= H_0/100$} km/s/Mpc. Additionally, we present the marginalized
one-dimensional posterior
distribution, where the dashed curve stands for the average likelihood distribution.}}
}
\end{figure*}

Let us now proceed to constrain the free parameter of the model  using two
different data sets, namely $CC+H_0$ and the combination of all data sets  $CC$ + $H_0$ +
SNeIa + BAO, following the procedure described in the previous section. In Figs.
\ref{figfT1_cc}  and \ref{figfT1_joint} we present the contour plots of various quantities
for $CC+H_0$ and  $CC$ + $H_0$ + SNeIa + BAO, respectively. Additionally, in Tables
\ref{f1-cc} and \ref{f1-joint} we summarize the best fit values for the two data sets
respectively.
\begin{table*}[ht]
\begin{center}
                \begin{tabular}{ccccc}
          \hline
          \hline
Parameter & best-fit & mean$\pm$ 1$\sigma$ & 95\% lower & 95\% upper \\ \hline
$\Omega_{cdm0 }$ &$0.1791$ & $0.1812_{-0.019}^{+0.016}$ & $0.1484$ & $0.2152$ \\
$h$ &$0.7303$ & $0.7285_{-0.018}^{+0.017}$ & $0.6946$ & $0.7623$ \\
$b$ &$0.03639$ & $0.03329_{-0.035}^{+0.043}$ & $-0.04738$ & $0.1096$ \\
$\Omega_{m0 }$ &$0.2291$ & $0.2312_{-0.019}^{+0.016}$ & $0.1984$ & $0.2652$ \\
          \hline
          \hline
          \end{tabular}
          \caption{	\label{f1-cc}
          Summary of the best fit values and main results for various
quantities and for the free parameter $b$,  for the  $f_{1}$CDM  model: $f(T)=\alpha
(-T)^{b}$ of
(\ref{modf1}), using   $CC$ + $H_0$ observational data. The parameter $\Omega_m$ includes
both baryons and cold dark matter,
i.e. $\Omega_m= \Omega_{cdm}+ \Omega_b$,
and \textbf{$h= H_0/100$} km/s/Mpc. }
      \end{center}
\end{table*}
\begin{table*}[ht]
	\begin{center}
		\begin{tabular}{ccccc}
			\hline
			\hline
			Parameter & best-fit & mean$\pm$ 1$\sigma$ & 95\% lower & 95\%
upper
\\ \hline
			$\Omega_{cdm0 }$ &$0.1806$ & $0.1835_{-0.019}^{+0.016}$ &
$0.1503$
& $0.2179$ \\
			$h$ &$0.7292$ & $0.7275_{-0.018}^{+0.017}$ & $0.6945$ & $0.7616$
\\
			$b$ &$0.05536$ & $0.05128_{-0.019}^{+0.025}$ & $0.00622$ &
$0.09329$ \\
			$\Omega_{m 0}$ &$0.2306$ & $0.2335_{-0.019}^{+0.016}$ & $0.2003$
&
$0.2679$ \\
			\hline
			\hline
		\end{tabular}
		\caption{
	\label{f1-joint}Summary of the best fit values and main results for various
quantities and for the free parameter $b$,  for the  $f_{1}$CDM  model: $f(T)=\alpha
(-T)^{b}$ of
(\ref{modf1}), using   $CC$ + $H_0$+ SNeIa + BAO observational data. The parameter
$\Omega_m$ includes
both baryons and cold dark matter,
i.e. $\Omega_m= \Omega_{cdm}+ \Omega_b$,
and \textbf{$h= H_0/100$} km/s/Mpc.}
\end{center}
\end{table*}

As we can see, the parameter $b$ which determines the deviation  from the $\Lambda$CDM
scenario is near zero, i.e. the $f(T)$ model at hand constrained by the current 
data is 
found to be
close to $\Lambda$CDM
cosmology as expected.  However, it is interesting to note that the $b=0$ value 
is
only marginally allowed, and hence the observations seem to slightly favor a small but 
not  non-zero deviation from  $\Lambda$CDM cosmology. This is one of 
the main results of the present work.

\subsection {$f_{2}$CDM  model:
$f(T)=\alpha T_{0}(1-e^{-p\sqrt{T/T_{0}}})$}

\begin{figure}[ht]
\centering
\includegraphics[width=0.63\linewidth]{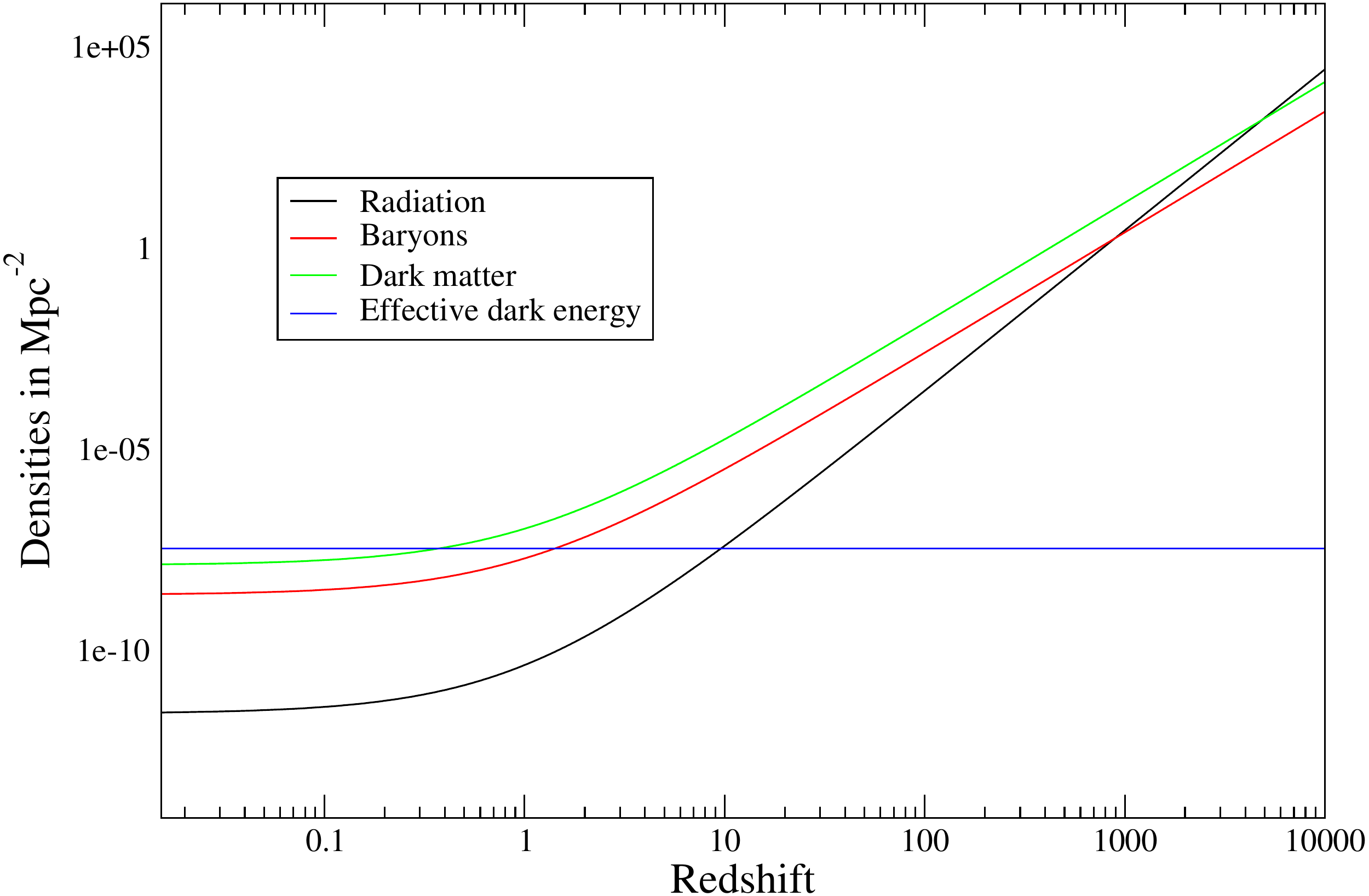}
\caption{{\it{
The evolution of various densities in units of $Mpc^{-2}$, multiplied by $8
\pi G/3$, as a function of the redshift, for the  $f_{2}$CDM  model: $f(T)=\alpha
T_{0}(1-e^{-p\sqrt{T/T_{0}}})$ of
(\ref{modf2}).
 We have considered $b = 1/p =0.1$, $h = 0.68$, $\Omega_{b0} = 0.05$, $\Omega_{cdm0} =
0.24$,
 $\Omega_{r0} = 10^{-5}$. }}}
\label{fT1_2}
\end{figure}

For the case of $f_{2}$CDM  model: $f(T)=\alpha T_{0}(1-e^{-p\sqrt{T/T_{0}}})$ of
(\ref{modf2}), in Fig.
\ref{fT1_2}   we   depict the evolution of the various
densities   between the
redshift range $z \in$ [0,10000]. Similarly to the previous model,  the density evolutions
are consistent with the thermal history of the
universe, i.e. we obtain   successively the radiation era, the matter era, and at
recent times (the transition is around $z \sim 0.70$) the onset of the dark-energy era
and of cosmic acceleration.

\begin{figure*}
  \includegraphics[width=5.0in, height=5.0in]{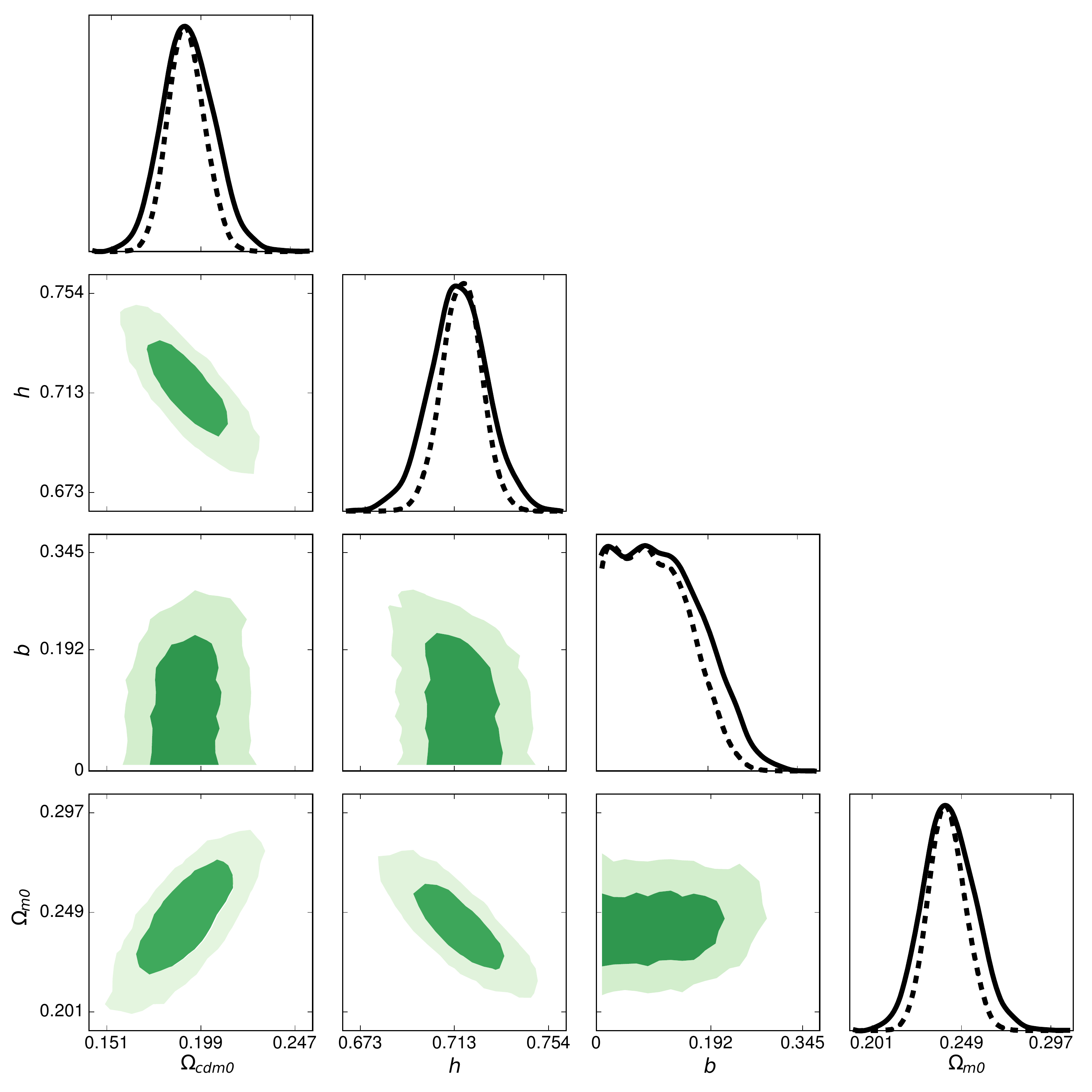}
  \caption{\label{figfT2_cc}
 { \it{ 68.3$\%$ and 95.4$\%$ confidence-level contour plots for various quantities and
for the free parameter $b$, for the  $f_{2}$CDM  model: $f(T)=\alpha
T_{0}(1-e^{-p\sqrt{T/T_{0}}})$ of
(\ref{modf2}), using only $CC$ + $H_0$
observational data. The parameter $\Omega_m$ includes both baryons and cold dark matter,
i.e. $\Omega_m= \Omega_{cdm}+ \Omega_b$,
and \textbf{$h= H_0/100$} km/s/Mpc. Additionally, we present the marginalized
one-dimensional posterior
distribution, where the dashed curve stands for the average likelihood distribution.}}}
\end{figure*}

\begin{figure*}
	\includegraphics[width=5.0in, height=5.0in]{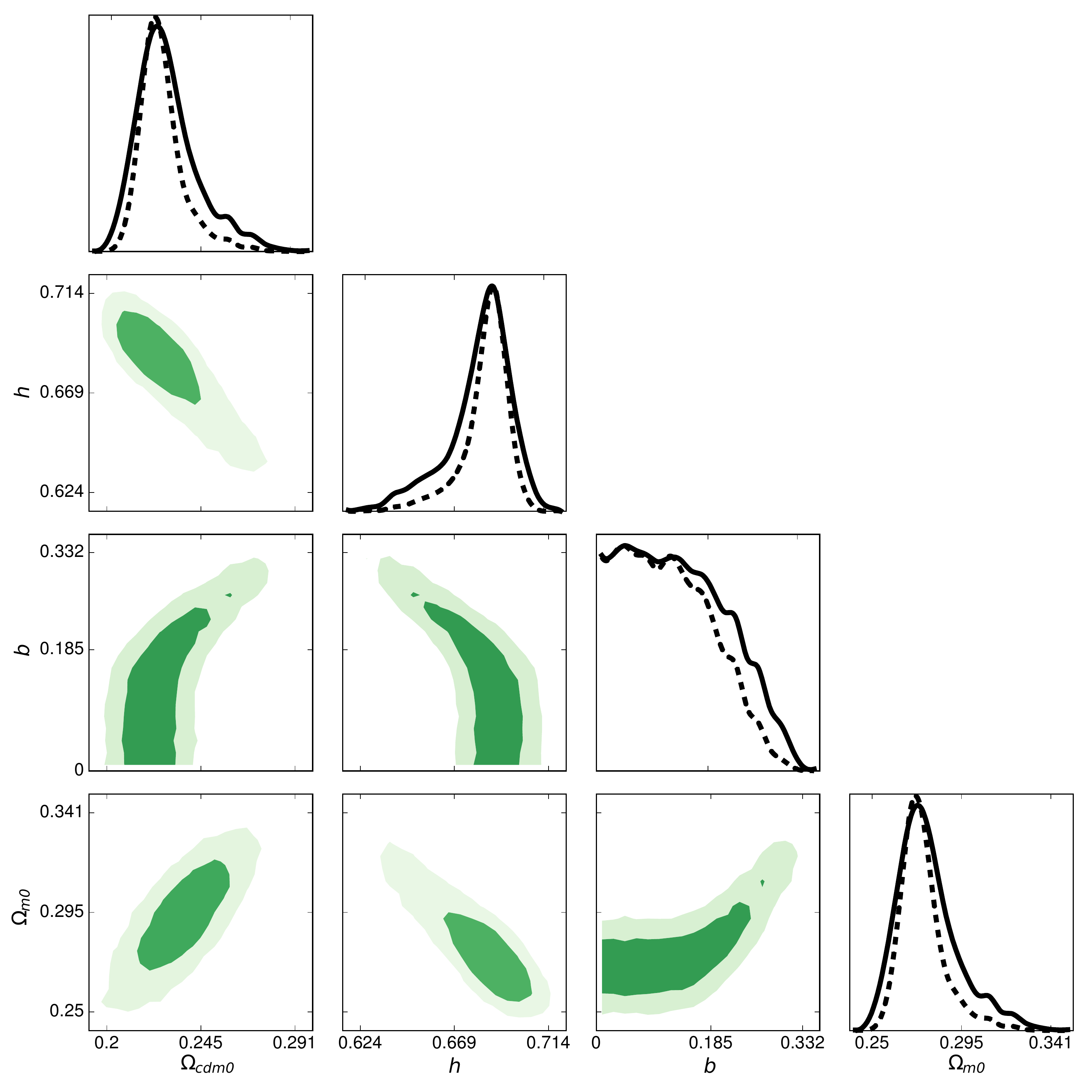}
	\caption{
	 { \it{ 68.3$\%$ and 95.4$\%$ confidence-level contour plots for various
quantities and
for the free parameter $b$, for the  $f_{2}$CDM  model: $f(T)=\alpha
T_{0}(1-e^{-p\sqrt{T/T_{0}}})$ of
(\ref{modf2}), using   $CC$ + $H_0$ + SNeIa + BAO
observational data. The parameter $\Omega_m$ includes both baryons and cold dark matter,
i.e. $\Omega_m= \Omega_{cdm}+ \Omega_b$,
and \textbf{$h= H_0/100$} km/s/Mpc. Additionally, we present the marginalized
one-dimensional posterior
distribution, where the dashed curve stands for the average likelihood distribution.}}
}
\label{figfT2_joint}
\end{figure*}

Let us now proceed to constrain this model  using two different data sets, namely $CC+H_0$
and the combination of all data sets  $CC$ + $H_0$ + SNeIa + BAO. In Figs.
\ref{figfT2_cc}  and \ref{figfT2_joint} we depict the contour plots of various
quantities for $CC+H_0$ and  $CC$ + $H_0$ + SNeIa + BAO, respectively. Additionally, in
Tables \ref{f2-cc} and \ref{f2-joint} we summarize the best fit values for the two data
sets respectively.

As we can see, the parameter $b$ which determines the deviation  from the $\Lambda$CDM
scenario is near zero,  i.e. the $f(T)$ model at hand  constrained by the current 
data is 
found to be
close to $\Lambda$CDM
cosmology as expected. However, contrary to the $f_{1}$CDM model of the previous 
subsection,   and due to its different functional form, it is interesting to 
note that the $b=0$ value is now inside the 1$\sigma$ region, and hence this model can 
observationally coincide with $\Lambda$CDM cosmology.

\begin{table*}[ht]
      \begin{center}
         \begin{tabular}{ccccc}
          \hline
          \hline
Parameter & best-fit & mean$\pm$ 1$\sigma$ & 95\% lower & 95\% upper \\ \hline
$\Omega_{cdm0 }$ &$0.1906$ & $0.1928_{-0.015}^{+0.014}$ & $0.1646$ & $0.2216$ \\
$h$ &$0.7179$ & $0.7153_{-0.013}^{+0.013}$ & $0.6884$ & $0.7429$ \\
$b$ &$0.04334$ & $0.1115_{-0.11}^{+0.035}$ & $1.68e-05$ & $0.2372$ \\
$\Omega_{m0 }$ &$0.2406$ & $0.2428_{-0.015}^{+0.014}$ & $0.2146$ & $0.2716$ \\
          \hline
          \hline
          \end{tabular}\caption{	\label{f2-cc}
           Summary of the best fit values and main results for various
quantities and for the free parameter $b$,  for the  $f_{2}$CDM  model: $f(T)=\alpha
T_{0}(1-e^{-p\sqrt{T/T_{0}}})$ of
(\ref{modf2}), using   $CC$ + $H_0$ observational data. The parameter $\Omega_m$ includes
both baryons and cold dark matter,
i.e. $\Omega_m= \Omega_{cdm}+ \Omega_b$,
and \textbf{$h= H_0/100$} km/s/Mpc.}
      \end{center}
\end{table*}
\begin{table*}[!]
      \begin{center}
     \begin{tabular}{ccccc}
          \hline
          \hline
Parameter & best-fit & mean$\pm$ 1$\sigma$ & 95\% lower & 95\% upper \\ \hline
$\Omega_{cd0m }$ &$0.2198$ & $0.2284_{-0.019}^{+0.0097}$ & $0.1996$ & $0.2641$ \\
$h$ &$0.6909$ & $0.6819_{-0.0093}^{+0.019}$ & $0.6435$ & $0.7103$ \\
$b$ &$0.04095$ & $0.1325_{-0.13}^{+0.043}$ & $1.403e-05$ & $0.2785$ \\
$\Omega_{m 0}$ &$0.2698$ & $0.2784_{-0.019}^{+0.0097}$ & $0.2496$ & $0.3141$ \\
          \hline
          \hline
          \end{tabular}
          \caption{ 	\label{f2-joint}Summary of the best fit values and main results
for various
quantities and for the free parameter $b$,  for the  $f_{2}$CDM  model: $f(T)=\alpha
T_{0}(1-e^{-p\sqrt{T/T_{0}}})$ of
(\ref{modf2}), using   $CC$ + $H_0$+ SNeIa + BAO observational data. The parameter
$\Omega_m$ includes
both baryons and cold dark matter,
i.e. $\Omega_m= \Omega_{cdm}+ \Omega_b$,
and \textbf{$h= H_0/100$} km/s/Mpc.}
      \end{center}
\end{table*}

 \subsection {$f_{3}$CDM  model:
$f(T)=\alpha T_{0}(1-e^{-pT/T_{0}})$}

\begin{figure}[ht]
\centering
\includegraphics[width=0.63\linewidth]{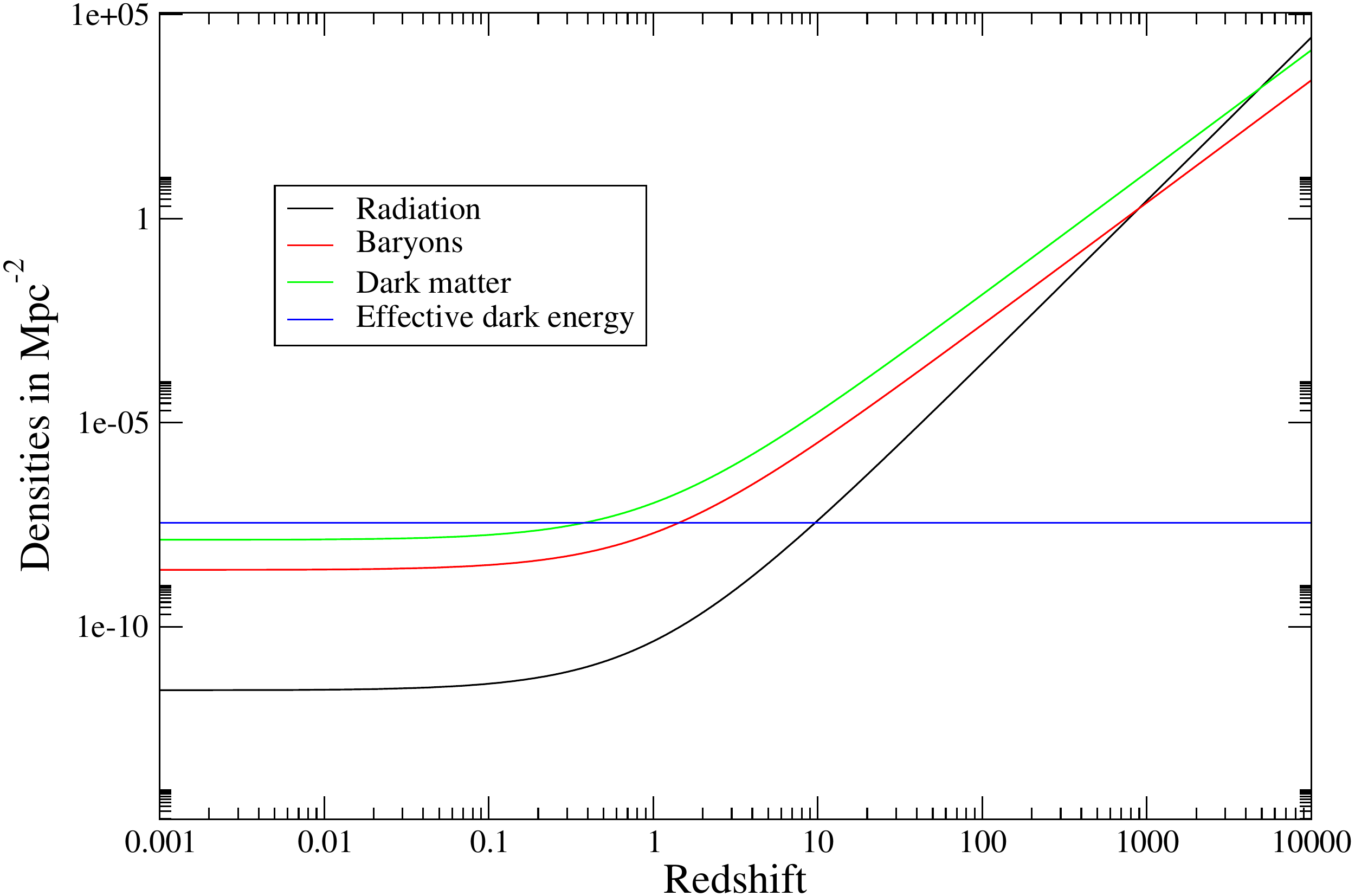}
\caption{{\it{The evolution of various densities in units of $Mpc^{-2}$, multiplied by $8
\pi G/3$, as a function of the redshift, for the $f_3$CDM model:  $f(T)=\alpha
T_{0}(1-e^{-pT/T_{0}})$ of
(\ref{modf3}).
 We have considered $b=1/p = 0.1$, $h = 0.68$, $\Omega_{b0} = 0.05$, $\Omega_{cdm0} =
0.24$,
   $\Omega_{r0} = 10^{-5}$.}}}
\label{fT1_3}
\end{figure}

For the case of $f_{3}$CDM  model: $f(T)=\alpha T_{0}(1-e^{-pT/T_{0}})$ of
(\ref{modf3}), in Fig. \ref{fT1_3}   we   depict the evolution of the various
densities   between the
redshift range $z \in$ [0,10000]. Similarly to the previous models,  the density
evolutions
are consistent with the thermal history of the
universe, i.e. we obtain   successively the radiation era, the matter era, and at
recent times (the transition is around $z \sim 0.70$) the onset of the dark-energy era
and of cosmic acceleration.

\begin{figure*}
  \includegraphics[width=5.0in, height=5.0in]{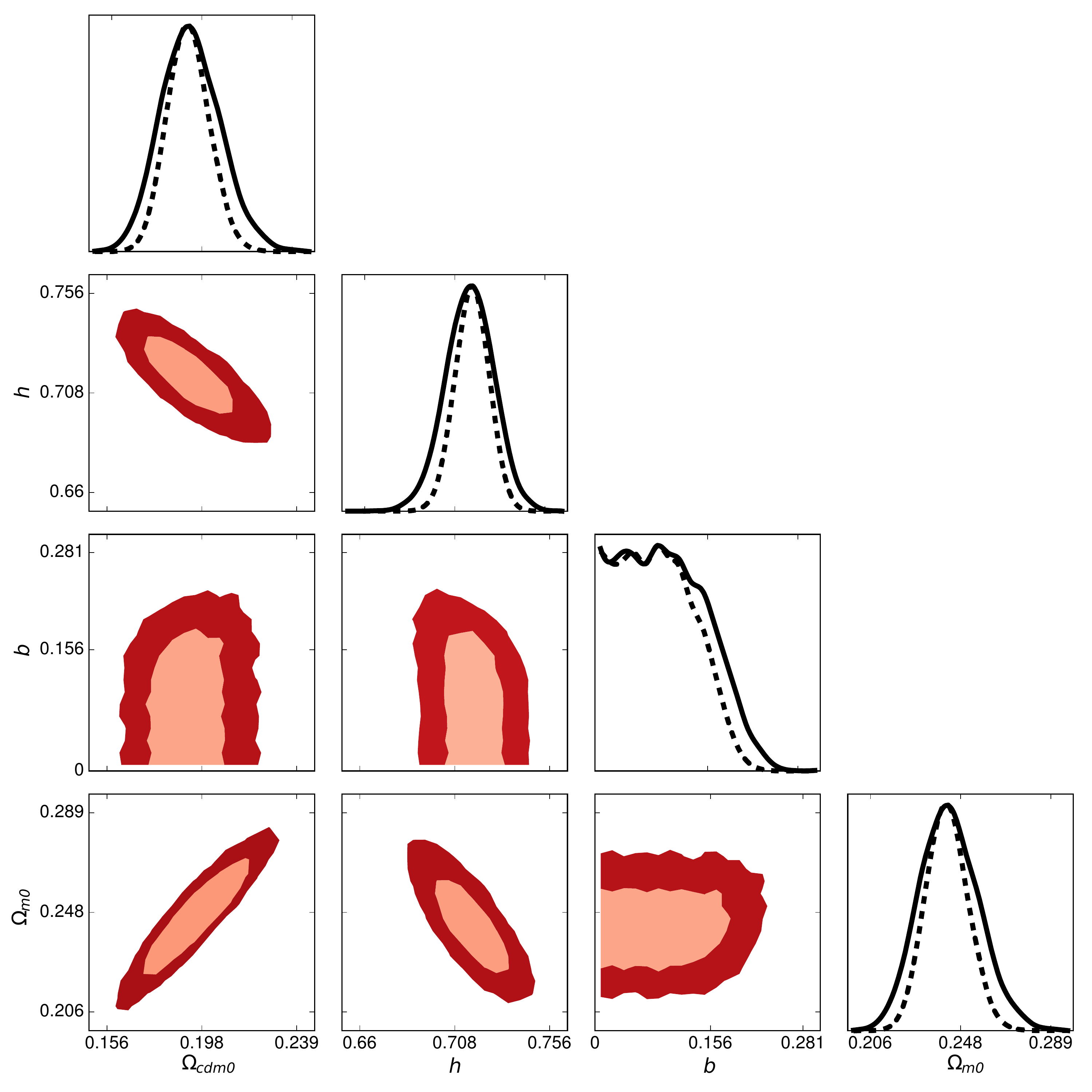}
  \caption{\label{figfT3_cc}
 { \it{ 68.3$\%$ and 95.4$\%$ confidence-level contour plots for various quantities and
for the free parameter $b$, for the  $f_{3}$CDM:  $f(T)=\alpha T_{0}(1-e^{-pT/T_{0}})$ of
 (\ref{modf3}), using only $CC$ + $H_0$
observational data. The parameter $\Omega_m$ includes both baryons and cold dark matter,
i.e. $\Omega_m= \Omega_{cdm}+ \Omega_b$,
and \textbf{$h= H_0/100$} km/s/Mpc. Additionally, we present the marginalized
one-dimensional posterior
distribution, where the dashed curve stands for the average likelihood distribution.}}}
\end{figure*}

\begin{figure*}
	\includegraphics[width=5.0in, height=5.0in]{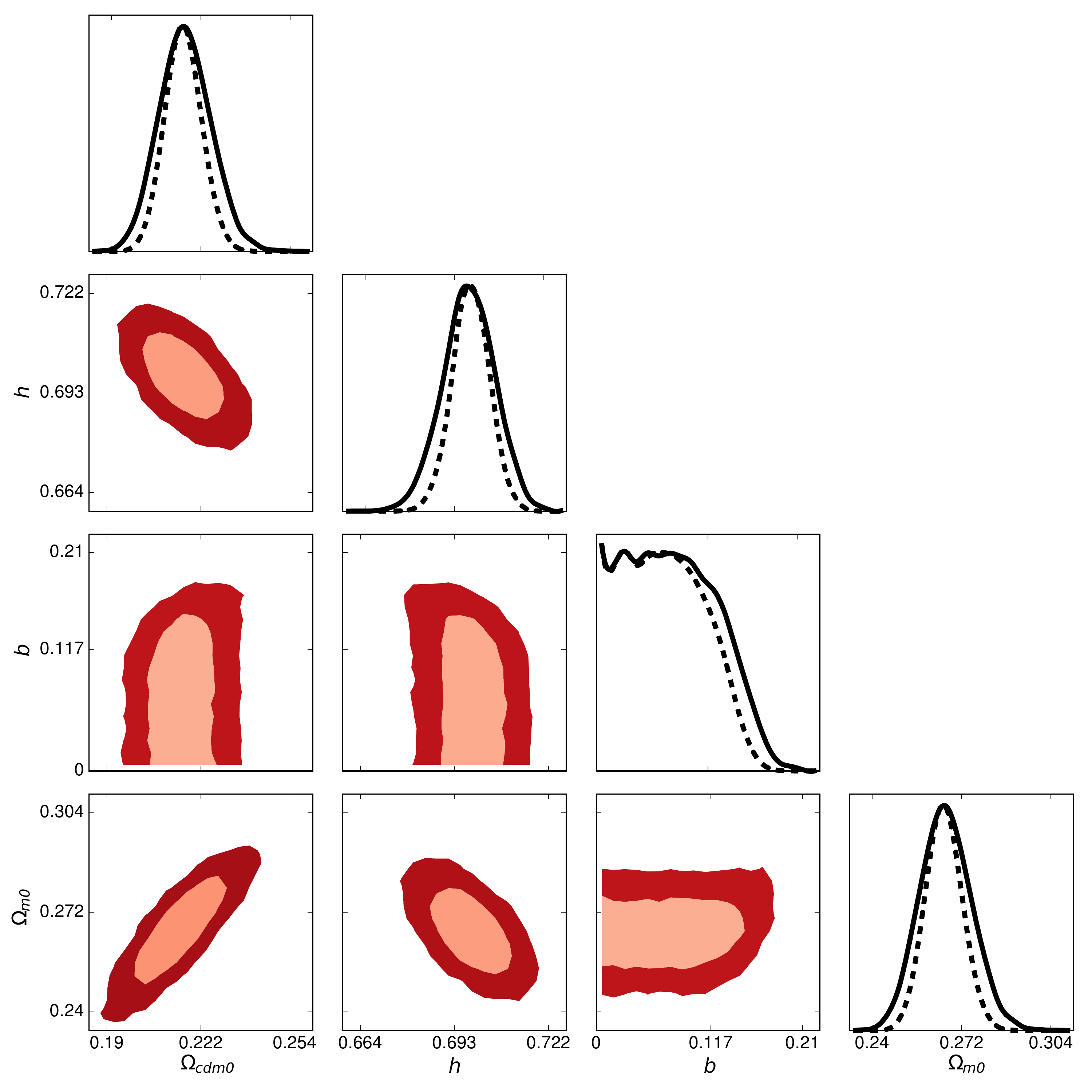}
	\caption{
	 { \it{ 68.3$\%$ and 95.4$\%$ confidence-level contour plots for various
quantities and
for the free parameter $b$, for the  $f_{3}$CDM:  $f(T)=\alpha T_{0}(1-e^{-pT/T_{0}})$ of
 (\ref{modf3}), using   $CC$ + $H_0$ + SNeIa + BAO
observational data. The parameter $\Omega_m$ includes both baryons and cold dark matter,
i.e. $\Omega_m= \Omega_{cdm}+ \Omega_b$,
and \textbf{$h= H_0/100$} km/s/Mpc. Additionally, we present the marginalized
one-dimensional posterior
distribution, where the dashed curve stands for the average likelihood distribution.}}
}
\label{figfT3_joint}
\end{figure*}

Let us now proceed to constrain this model  using two different data sets, namely
$CC+H_0$
and the combination of all data sets  $CC$ + $H_0$ + SNeIa + BAO. In Figs.
\ref{figfT3_cc}  and \ref{figfT3_joint} we depict the contour plots of various
quantities for $CC+H_0$ and  $CC$ + $H_0$ + SNeIa + BAO, respectively. Additionally, in
Tables \ref{f3-cc} and \ref{f3-joint} we summarize the best fit values for the two data
sets respectively.

As we can see, the parameter $b$ which determines the deviation  from the $\Lambda$CDM
scenario is near zero, i.e. the $f(T)$ model at hand  constrained by the current 
data is 
found to be
close to $\Lambda$CDM
cosmology as expected. However, similarly to $f_{2}$CDM model, and contrary to the 
$f_{1}$CDM  model, it is interesting to note that the $b=0$ value is now inside the
1$\sigma$ region, and hence this model can observationally coincide with $\Lambda$CDM
cosmology.  The similarities in the behavior of $f_{2}$CDM and $f_{3}$CDM models 
arise from their similar functional form, and as we can see  in a cosmological context 
they are practically and statistically indistinguishable.


\begin{table*}[ht]
      \begin{center}
       \begin{tabular}{ccccc}
 \hline
          \hline
Parameter & best-fit & mean$\pm\sigma$ & 95\% lower & 95\% upper \\ \hline
$\Omega_{cdm0 }$ &$0.1902$ & $0.1924_{-0.015}^{+0.013}$ & $0.1647$ & $0.2209$ \\
$h$ &$0.7184$ & $0.7157_{-0.013}^{+0.013}$ & $0.689$ & $0.7417$ \\
$b$ &$0.04622$ & $0.106_{-0.09}^{+0.052}$ & $1.228e-05$ & $0.2246$ \\
$\Omega_{m 0}$ &$0.2402$ & $0.2424_{-0.015}^{+0.013}$ & $0.2147$ & $0.2709$ \\
 \hline
          \hline
 \end{tabular}
 \caption{	\label{f3-cc}
           Summary of the best fit values and main results for various
quantities and for the free parameter $b$,  for the  $f_{3}$CDM:  $f(T)=\alpha
T_{0}(1-e^{-pT/T_{0}})$ of
 (\ref{modf3}), using   $CC$ + $H_0$ observational data. The parameter $\Omega_m$
includes
both baryons and cold dark matter,
i.e. $\Omega_m= \Omega_{cdm}+ \Omega_b$,
and \textbf{$h= H_0/100$} km/s/Mpc.}
      \end{center}
\end{table*}

\begin{table*}[!]
      \begin{center}
     \begin{tabular}{ccccc}
 \hline
          \hline
Parameter & best-fit & mean$\pm\sigma$ & 95\% lower & 95\% upper \\ \hline
$\Omega_{cdm0 }$ &$0.2148$ & $0.2162_{-0.0099}^{+0.0091}$ & $0.1973$ & $0.235$ \\
$h$ &$0.6993$ & $0.698_{-0.0084}^{+0.0089}$ & $0.6807$ & $0.7153$ \\
$b$ &$0.03207$ & $0.09_{-0.08}^{+0.041}$  & $1.483e-05$ & $0.153$ \\
$\Omega_{m0}$ &$0.2648$ & $0.2662_{-0.0099}^{+0.0091}$ & $0.2473$ & $0.285$ \\
 \hline
          \hline
 \end{tabular}
          \caption{ 	\label{f3-joint}Summary of the best fit values and main results
for various
quantities and for the free parameter $b$,  for the  $f_{3}$CDM:  $f(T)=\alpha
T_{0}(1-e^{-pT/T_{0}})$ of
 (\ref{modf3}), using   $CC$ + $H_0$+ SNeIa + BAO observational data. The parameter
$\Omega_m$ includes
both baryons and cold dark matter,
i.e. $\Omega_m= \Omega_{cdm}+ \Omega_b$,
and \textbf{$h= H_0/100$} km/s/Mpc.}
      \end{center}
\end{table*}

\section{Conclusions}
\label{Conclusions}

In the present work we have used the  recently released cosmic chronometers data in order
to impose constraints on the viable and most used $f(T)$ gravity models. In particular,
we considered three $f(T)$ models with two parameters, out of which one is
independent, that are known to pass the basic observational tests, and we
quantified their deviation from $\Lambda$CDM cosmology through a sole parameter.  Hence,
we used observational data in order to fit  this parameter,  as well as various other
cosmological quantities.

In our investigation we used (i) the very recently released cosmic chronometer data sets
along with the latest measured value of the local Hubble parameter, $H_0 = 73.02 \pm
1.79$ km/s/Mpc \cite{riess}, (ii)  the Union 2.1 compilation \cite{snia3}, which contains
580 SNeIa data in the redshift range $0.015 \leq z \leq 1.41$, (iii) baryon acoustic
oscillation data points. For each specific $f(T)$ model we provided two different sets of
values for the model parameters, one  arising from 
the combined data 
$CC$ + $H_0$,  
and one arising from   
$CC$ + $H_0$  + SNeIa + BAO data.

As we saw, for the first model, namely $f_{1}$CDM one, both  $CC$ + $H_0$ as well as $CC$
+ $H_0$  + SNeIa + BAO analysis seem to slightly favor a small but non-zero deviation from
$\Lambda$CDM cosmology. On the other hand, for $f_{2}$CDM and $f_{3}$CDM models,
deviation from $\Lambda$CDM cosmology is also allowed, nevertheless the best-fit value is
very close to its $\Lambda$CDM one.  These results are in qualitative agreement 
with previous observational studies on $f(T)$ gravity, where the CMB shift parameter 
alongside with  SNeIa + BAO data had been used 
\cite{Wu:2010mn,Nesseris:2013jea,Capozziello:2015rda,Basilakos:2016xob}, however the 
incorporation of the novel cosmic chronometer data, as well as the improved BAO ones, has 
led to smaller statistical errors. Hence, although in the previous analyses all viable 
$f(T)$ models in the obtained confidence level were found to be practically and 
statistically indistinguishable from $\Lambda$CDM cosmology,   in the present work we 
found 
that one of them, namely the power-law model $f_{1}$CDM model of \cite{Bengochea:2008gz}, 
may have a slight deviation from $\Lambda$CDM paradigm.

In summary, using for the first time the recently released cosmic chronometer 
data, we
were able to fit the viable and most used  $f(T)$ gravity models. Clearly, $f(T)$ gravity
is consistent with observations, and hence it can still serve as a candidate for modified
gravity.  It would be interesting to extend the present analysis by including  
the 
anisotropy in the cosmic microwave background (CMB), data from large scale structure 
(LSS), as 
well as data from weak lensing (WL), in order to obtain  more complete results and acquire
  a clearer picture concerning the possible deviation from  $\Lambda$CDM cosmology. 
Such a full combined analysis using dynamical tests is left for a future project.

 \begin{acknowledgments}
SP acknowledges the Science and Engineering Research Board (SERB), Govt. of India, for
awarding National Post-Doctoral Fellowship (File No: PDF/2015/000640).
This article is based upon work from COST Action ``Cosmology and Astrophysics Network
for Theoretical Advances and Training Actions'', supported by COST (European Cooperation
in Science and Technology).  Finally, the authors would like to thank an anonymous 
referee for his/her valuable suggestions. 
 \end{acknowledgments}

\end{document}